\documentclass[sigconf]{acmart}

\renewcommand\footnotetextcopyrightpermission[1]{} 
\usepackage{epsfig,endnotes,url,color,subfigure}
\usepackage{array}
\usepackage{multirow}
\usepackage{xspace}
\usepackage{verbatim}
\usepackage{wrapfig}
\usepackage[english]{babel}
\usepackage{amsmath}
\usepackage{caption}
\usepackage{balance}
\usepackage[ruled, vlined, linesnumbered]{algorithm2e}

\usepackage{tcolorbox}

\usepackage{hyperref}
\hypersetup{pdftex,colorlinks=true,allcolors=blue}
\usepackage{hypcap}
\usepackage{xr}

\newcommand{\projecttitle}{\textsc{ApproxJoin}\xspace}
\newcommand{\myfontsize}{\fontsize{8}{9}\selectfont}
\newcommand{\commentfontsize}{\fontsize{7}{8}\selectfont}

\makeatletter
\def\@copyrightspace{\relax}
\makeatother
\setcopyright{none}

\usepackage{grumble}

\newcommand{\myparagraph}[1]{\smallskip \noindent{\bf {#1}.}}

\newcommand{\out}[1] {}

\newcounter{codeLineCntr}

\reversemarginpar
\newif\ifnotes
\notestrue

\newcommand{\punt}[1]{}

\renewcommand{\eqref}[1]{Equation~(\ref{eq:#1})}

\newcommand{\proc}[1]{\ifmmode\mbox{\textsc{#1}}\else\textsc{#1}\fi}

  \newcommand{\func}[1]{\ifmmode\mathrm{#1}\else\textrm{#1}fi} %

\newcounter{remark}[section]

\usepackage[inline]{enumitem}
\setlist{noitemsep,topsep=0pt,parsep=0pt,partopsep=0pt}

\usepackage{setspace}
\usepackage[margin=5pt,font={stretch=0.9}]{caption}

\begin{document}
\title{Approximate Distributed Joins in Apache Spark}

\author{Do Le Quoc$^\dag$, 
Istemi Ekin Akkus$^\ddag$,\\ Pramod Bhatotia$^\star$,  Spyros Blanas$^\#$,  Ruichuan Chen$^\ddag$, \\ Christof Fetzer$^\dag$, Thorsten Strufe$^\dag$\\
\small {$^\dag$TU Dresden, Germany  \quad $^\ddag$Nokia Bell Labs, Germany \quad $^\star$University of Edinburgh, UK \quad $^\#$The Ohio State University, USA} \\
{\small Technical Report, May 2018}
}

\begin{abstract}

The join operation is a fundamental building block of parallel data processing. Unfortunately, it is very resource-intensive to compute an equi-join across massive datasets. The approximate computing paradigm allows users to trade accuracy and latency for expensive data processing operations. The equi-join operator is thus a natural candidate for optimization using approximation techniques. Although sampling-based approaches are widely used for approximation, sampling over joins is a compelling but challenging task regarding the output quality. Naive approaches, which perform joins over dataset samples,  would not preserve statistical properties of the join output.

To realize this potential, we interweave Bloom filter sketching and stratified sampling with the join computation in a new operator, \projecttitle, that preserves the statistical properties of the join output. \projecttitle leverages a Bloom filter to avoid shuffling non-joinable data items around the network and then applies stratified sampling to obtain a representative sample of the join output. 

Our analysis shows that \projecttitle scales well and significantly reduces data movement, without sacrificing tight error bounds on the accuracy of the final results. 
We implemented \projecttitle in Apache Spark and evaluated \projecttitle using microbenchmarks and real-world case studies. The evaluation shows that \projecttitle achieves a speedup of $6-9\times$ over unmodified Spark-based joins with the same sampling rate. Furthermore, the speedup is accompanied by a significant reduction in the shuffled data volume, which is $5-82\times$ less than unmodified Spark-based joins.

\end{abstract}

\maketitle

 \section{Introduction}
\label{sec:introduction}

The volume of digital data has grown exponentially over the last ten years.
A key contributor to this growth has been loosely-structured raw data that 
are perceived to be cost-prohibitive to clean, organize and store in a
database management system (DBMS).
These datasets are frequently stored in data repositories (often
called ``data lakes'') for just-in-time querying and analytics.
Extracting useful knowledge from a data lake is a challenge since it
requires data analytics systems that 
adapt to variety in the output of different data sources
and 
answer ad-hoc queries over vast amounts of data quickly.

To pluck the valuable information from raw data, data processing 
frameworks such as Hadoop~\cite{hadoop}, Apache Spark~\cite{spark} and Apache
Flink~\cite{flink} are widely used to perform ad-hoc data manipulations and
then combine data from different input sources using a \emph{join operation}.
While joins are a critical building block of any analytics pipeline, they are
expensive to perform, especially with regard to communication costs in
distributed settings. It is not uncommon for a parallel data processing
framework to take hours to process a complex join query~\cite{Hive2010}.

\begin{figure}[t]
	\centering
	\includegraphics [scale=0.55]{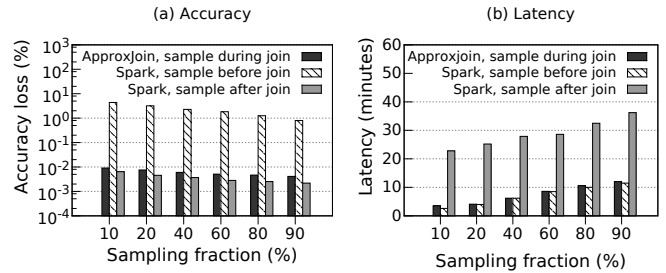} 
    \caption{Comparison between different sampling strategies for distributed join with varying sampling fractions. }
	\label{fig:sampling-join-comparison}
\end{figure}

Parallel data processing frameworks are thus embracing \emph{approximate computing}
to answer complex queries over massive datasets quickly 
\cite{BlinkDB,BlinkDB-2,quickr-sigmod,wander-join}.
The approximate computing paradigm is based on the observation that approximate rather than
exact results suffice if real-world applications can reason about measures of
statistical uncertainty such as confidence intervals~\cite{approx-ex-1,approx-ex-2}.
Such applications sacrifice accuracy for lower latency by processing only a
fraction of massive datasets.
What response time and accuracy targets are acceptable for each particular
problem is determined by the user that has the necessary domain expertise.

However, approximating join results by sampling
is an inherently difficult problem from a correctness perspective, because
uniform random samples of the join inputs cannot construct an unbiased
random sample of the join output \cite{samplingoverjoin}.
In practice, as shown in Figure~\ref{fig:sampling-join-comparison}, 
sampling input datasets before the join and joining the samples sacrifices up 
to an order of magnitude in accuracy; sampling after the join is
accurate but also $3 - 7\times$ slower due to the data that are shuffled
to compute the join result.

Obtaining a \emph{correct and precondition-free sample} of the join output 
in a distributed computing framework is a challenging task.
Previous work has assumed some prior
knowledge about the joined tables, often in the form of an offline 
sample or a histogram~\cite{quickr-sigmod, BlinkDB, aqua}. 
Continuously maintaining histograms or samples over the entire
dataset ---PB of data--- is unrealistic as ad-hoc analytical queries 
process raw data selectively.
Join approximation techniques for a DBMS, like  
RippleJoin \cite{ripple-join} and WanderJoin \cite{wander-join},
have not considered the intricacies of HDFS-based processing where
random disk accesses are notoriously inefficient and data have not been
indexed in advance. In addition, 
both algorithms are designed for single-node join processing; parallelizing the optimization procedure for a Spark cluster is non-trivial.

In this work, we design a novel approximate distributed join 
algorithm that combines a {\em Bloom filter} sketching technique with {\em
stratified sampling} during the join operation. 
The Bloom filter curtails redundant shuffling of tuples that will not
participate in the subsequent join operations, thus reducing communication and
processing overhead.
In addition, \projecttitle automatically selects and progressively refines the
sampling rate to meet user-defined latency and quality requirements 
by using the Bloom filter to estimate the cardinality of the join output.

Once the sampling rate has been determined,
stratified sampling over the 
remaining tuples
produces
a sample of the join output that approximates the
result of the entire join.
However, sampling without coordination from concurrent processes can introduce
bias in the final result, adversely affecting the accuracy of the
approximation. 
\projecttitle removes this bias using the Central Limit Theorem
and the Horvitz-Thompson estimator.
The proposed join mechanism can be used for both {\em two-way} joins and {\em
multi-way joins}.
As shown in Figure~\ref{fig:sampling-join-comparison}, sampling during the 
join produces accurate results with fast response times.

We implemented \projecttitle in Apache
Spark~\cite{spark, spark-nsdi-2012} and evaluate its effectiveness via microbenchmarks, TPC-H
queries, and a real-world workload.
Our evaluation shows that \projecttitle achieves a speedup of $6 - 9\times$
over Spark-based joins with the same sampling fraction. 
\projecttitle leverages Bloom filtering to reduce the shuffled data volume
during the join operation by $5 - 82\times$ compared to Spark-based systems.
Without any sampling, our microbenchmark evaluation shows that \projecttitle 
achieves a speedup of $2 - 10\times$ over the native Spark
RDD join \cite{spark-nsdi-2012} and $1.06-3\times$ over a Spark
repartition join. 
In addition, our evaluation with TPC-H benchmark shows that \projecttitle
is $1.2 - 1.8\times$ faster than the state-the-art SnappyData system~\cite{snappydata}.

To summarize, our contributions are:
\begin{itemize}[leftmargin=1.5em]

	\item A novel mechanism to perform stratified sampling over joins in
		parallel computing frameworks that relies on a Bloom filter sketching
		technique to preserve the statistical quality of the join output.
	
	\item A progressive refinement procedure that automatically
		select a sampling rate that meets user-defined latency and accuracy
		targets for approximate join computation.
	
	\item An extensive evaluation of an implementation of \projecttitle in
		Apache Spark using microbenchmarks, TPC-H queries, and a real-world
		workload that shows that \projecttitle outperforms native Spark-based
		joins and the state-of-the-art SnappyData system by a substantial margin.

\end{itemize}

The remainder of the paper is organized as follows. We first provide an overview of the system ($\S$\ref{sec:overview}). Next, we describe the our approach to mitigate the overhead of distributed join operations ($\S$\ref{sec:design}).  Thereafter, we present implementation of \projecttitle ($\S$\ref{sec:implementation}), and evaluation ($\S$\ref{sec:evaluation} \& $\S$\ref{sec:case-studies}) of \projecttitle.
Finally, we present the related work and conclusions in $\S$\ref{sec:related} and $\S$\ref{sec:conclusion} respectively.

\section{Overview}
\label{sec:overview}

\projecttitle is designed to mitigate the overhead of distributed join operations in big data analytics systems, such as Apache Flink or Apache Spark.
The input of \projecttitle consists of several datasets to be joined. 
We facilitate joins on the input datasets by providing a simple user interface. The user submits the join query and its corresponding query execution budget. The query budget can be in the form of expected latency guarantees, or the desired accuracy level. Our system ensures that the input data is processed within the specified query budget.
 To achieve this, \projecttitle applies the approximate computing paradigm by processing only a partial input data items from the datasets to produce an approximate output with error bounds. At a high level, \projecttitle makes use of a combination of sketching and sampling to select a subset of input datasets based on the user specified query budget.

\myparagraph{Design goals} We had the following goals when we designed and implemented \projecttitle:

\begin{itemize}[leftmargin=1.5em]
	
	\item \textit{Transparency:} Provide a simple programming interface to users that is similar to the join operation of state-of-the-art systems. This goal implies that there will be negligible (or no) modifications to existing programs.
	
	\item \textit{Query budget guarantees:} Ensure that the join operation is performed within the query budget supplied by the user in the form of desired latency or desired error bound. This goal implies that the system should accurately estimate the latency and error bounds of the approximation in the join operation.
	
	\item \textit{Efficiency:} Handle very large input datasets in an efficient and cost-effective manner. This goal implies that the system reduces the usage of resources (e.g., network, CPU) as much as possible.
	
\end{itemize}

\myparagraph{Query interface}
\projecttitle supports joins with algebraic aggregation functions, such as SUM, AVG, COUNT, and STDEV. In addition, a query execution budget is provided to specify either the latency requirement or desired error bound. More specifically, 
 consider the case where a user wants to perform an aggregation query after an equal-join on attribute $A$ for $n$ input datasets $R_1 \Join R_2 \Join ... \Join R_n$, where $R_i (i = 1, ..., n)$ represents an  input dataset. The user sends the query $q$ and supplies a query budget $q_{budget}$ to \projecttitle. The query budget can be in the form of desired latency $d_{desired}$ or desired error bound $err_{desired}$. For instance, if the user wants to achieve a desired latency (e.g., $d_{desired} = 120$ seconds), or a desired error bound (e.g., $err_{desired} = 0.01$ with confidence level of $95\%$), he/she defines the query as follows:
\begin{tcolorbox}
	\bfseries
	{\myfontsize
	SELECT SUM($R_1.V$ + $R_2.V$ + ... + $R_n.V$)\\
	FROM $R_1$, $R_2$, ..., $R_n$\\
	WHERE $R_1.A$ = $R_2.A$ = ... = $R_n.A$\\
	WITHIN $120$ SECONDS\\
	\textit{OR}\\
	ERROR $0.01$ CONFIDENCE $95\%$}
	
\end{tcolorbox}

\projecttitle executes the query and returns the most accurate query result within the desired latency which is $120$ seconds, or returns the query result within the desired error bound $\pm 0.01$ at a $95$\% confidence level.

\begin{figure}[t]
\centering
\includegraphics[scale=0.3]{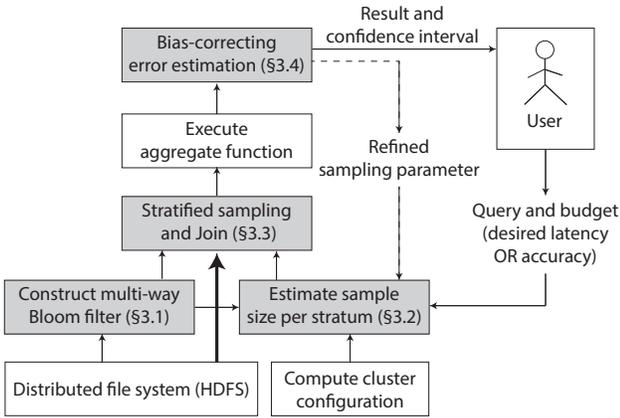}
\caption{ApproxJoin system overview (shaded boxes depict the implemented modules in Apache Spark).}
\label{fig:system-overview}
\end{figure}

\myparagraph{Design overview} 
The basic idea of \projecttitle is to address the shortcomings of the existing join operations in big data systems by reducing the number of data items that need to be processed. Our first intuition is that we can reduce the latency and computation of a distributed join by removing redundant transfer of data items that are not going to participate in the join. Our second intuition is that the exact results of the join operation may be desired, but not necessarily critical, so that an approximate result with well-defined error bounds can also suffice for the user.

Figure~\ref{fig:system-overview} shows an overview of our approach. There are two stages in \projecttitle for the execution of the user's query: 

\myparagraph{Stage \#1: Filtering redundant items} In the first stage, \projecttitle determines the data items that are going to participate in the join operation and filters the non-participating items. This filtering reduces the data transfer that needs to be performed over the network for the join operation. It also ensures that the join operation will not include `null' results in the output that will require special handling, as in WanderJoin \cite{wander-join}. \projecttitle employs a well-known data structure, {\em Bloom filter}~\cite{bloom-filter}. Our filtering algorithm executes in parallel at each node that stores partitions of the input and handles multiple input tables at the same time.

\myparagraph{Stage \#2: Approximation in distributed joins}
In the second stage, \projecttitle uses a sampling mechanism that is executed {\em during} the join process: we sample the input datasets while the cross product is being computed. This  mechanism overcomes the limitations of the previous approaches and enables us to achieve low latency as well as preserve the quality of the output as highlighted in Figure~\ref{fig:sampling-join-comparison}. Our sampling mechanism is executed during the join operation and preserves the statistical properties of the output. 

In addition, we combine our mechanism with {\em stratified sampling}~\cite{stratified-sampling}, where tuples with distinct join keys are sampled independently with simple random sampling. As a result, data items with different join keys are fairly selected to represent the sample, and no join key will be overlooked. The final sample will contain all join keys---even the ones with few data items---so that the statistical properties of the sample are preserved.

More specifically, \projecttitle executes the following steps for approximation in distributed joins:

\myparagraph{Step \#2.1: Determine sampling parameters}
\projecttitle employs a {\em cost function} to compute an optimal sample size according to the corresponding budget of the query. This computation ensures that the query is executed within the desired latency and error bound parameters of the user.

\myparagraph{Step \#2.2: Sample and execute query}
Using this sampling rate parameter, \projecttitle samples during the join and then executes the aggregation query $q$ using the obtained sample.

\myparagraph{Step \#2.3: Estimate error}
After executing the query, \projecttitle provides an approximate result together with a corresponding error bound in the form of $result \pm error\_bound$ to the user.

Note that our sampling parameter estimation provides an adaptive interface for selecting the sampling rate based on the user-defined accuracy and latency requirements. \projecttitle adapts by activating a feedback mechanism to refine the sampling rate after learning the data distribution of the input datasets (shown by the dashed line in Figure~\ref{fig:system-overview}).

\begin{figure}[t]
\centering
\includegraphics[scale=0.33]{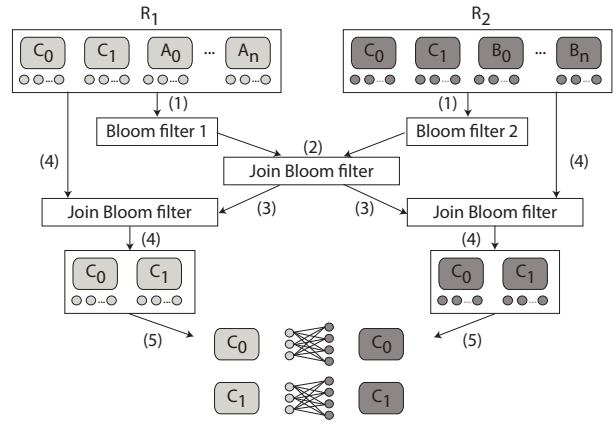}
\caption{Bloom filter building for two datasets. Algorithm~\ref{alg:subroutines-bloomfilter-joins} generalizes for distributed multi-way joins.}
\label{fig:approxjoin-bloomfilter}
\end{figure}

\section{Design}
\label{sec:design}

In this section, we explain the design details of \projecttitle. We first describe how we filter redundant data items for multiple datasets to support multiway joins (\S\ref{sec:bloom_multiway}). Then, we describe how we perform approximation in distributed joins using three main steps: (1) how we determine the sampling parameter to satisfy the user-specified query budget (\S\ref{sec:cost-function}), (2) how our novel sampling mechanism executes during the join operation (\S\ref{sec:sampling}), and finally (3) how we estimate the error for the approximation (\S\ref{sec:design-error-estimation}).

\subsection{Filtering Redundant Items}
\label{sec:bloom_multiway}

In a distributed setting, join operations can be expensive due to the communication cost of the data items. This cost can be especially high in multi-way joins, where several datasets are involved in the join operation.  One reason for this high cost is that data items not participating in the join are shuffled through the network during the join operation. 

To reduce this communication cost, we need to distinguish such redundant items and avoid transferring them over the network. In \projecttitle, we use Bloom filters for this purpose. The basic idea is to utilize Bloom filters as a compressed set of all items present at each node and combine them to find the intersection of the datasets used in the join. This intersection will represent the set of data items that are going to participate in the join.

A Bloom filter is a data structure designed to query the presence of an element in a dataset in a rapid and memory-efficient way~\cite{bloom-filter}. There are three advantages why we choose Bloom filters for our purpose. First, querying the membership of an element is efficient: it has $O(h)$ complexity, where $h$ denotes a constant number of hash functions. Second, the size of the filter is linearly correlated with the size of the input, but it is significantly smaller compared to the original input size. Finally, constructing a Bloom filter is fast and requires a single pass over the input.

Bloom filters have been exploited to improve distributed joins in the past~\cite{bloomfilter-join, bloomfilter-join1, bloomfilter-join2, hashjoin-bloomfilter}. However, these proposals support only two-way joins. Although one can cover joins with multiple input datasets by chaining two-way joins, this approach would add to the latency of the join results. \projecttitle handles multiple datasets at the same time and supports multi-way joins without introducing additional latency.

For simplicity, we first explain how our algorithm uses a Bloom filter to find the intersection of two input datasets. Afterwards, we explain how our algorithm finds the intersection of multiple datasets at the same time.

\myparagraph{I: Two-way Bloom filter} For the two-way filtering, consider the join operation of two datasets $R_1 \Join R_2$ (see Figure~\ref{fig:approxjoin-bloomfilter}). First, we construct a Bloom filter for each input (step 1 in Figure \ref{fig:approxjoin-bloomfilter}), which we refer to as {\em dataset filter}. We perform $AND$ among the dataset filters (step $2$). The resulting Bloom filter represents the intersection of both datasets and is referred to as {\em join filter}. 

Afterwards, we broadcast the join filter to all nodes (step $3$). Each node checks the membership of the data items in its respective input dataset in the join filter. If a data item is not present, it is discarded. In Figure \ref{fig:approxjoin-bloomfilter}, all data items with keys $C0$ and $C1$ are preserved.

\begin{algorithm}[t]

\myfontsize

\KwIn{
	{
		\\
		$n$: number of input datasets \\
		$|BF|$: size of the Bloom filter \\
		$fp$: false positive rate of the Bloom filter \\
		$R$: input datasets \\
	}
}
\BlankLine

{\commentfontsize // {\em Build a Bloom filter for the join input datasets $R$}}\\

\underline{{\bf buildJoinFilter}}($R$, $|BF|$, $fp$)\\
\Begin{
	{\commentfontsize // {\em Build a Bloom filter for each input $R_i$}}\\
        {\commentfontsize // {\em Executed in parallel at worker nodes}}\\
	$\forall i \in \{1... n\}$: BF$_i$ $\leftarrow$ {\tt buildInputFilter}($R_i$, $|BF|$, $fp$)\;
	{\commentfontsize // {\em Merge input filters BF$_i$ for the overlap between inputs}}\\
        {\commentfontsize // {\em Executed sequentially at master node}}\\
	BF $\leftarrow$ $\cap_{i=1}^{n}$BF$_i$\;
	\Return{BF} 
}

{\commentfontsize // {\em Build a Bloom filter for input $R_i$}}\\
{\commentfontsize // {\em Executed in parallel at worker nodes}}\\
\underline{{\bf buildInputFilter}}($R_i$, $|BF|$, $fp$)\\
\Begin{
	$|p_i|$ := number of partitions of input dataset $R_i$\\
	$p_i$ := $\{p_{i,j}\}$, where $j = 1, ..., |p_i|$\\
	{\commentfontsize //{\em MAP PHASE}}\\
    {\commentfontsize //{\em Initialize a filter for each partition}}\\
    \ForAll {$j$ in $\{1... |p_i|\}$} {
    	 p-BF$_{i,j}$ $\leftarrow$ {\tt BloomFilter}($|BF|$, $fp$)\; 
         $\forall r_j \in p_{i,j}$: p-BF$_{i,j}$.{\tt add}($r_j.key$)\;
     }     
    {\commentfontsize //{\em REDUCE PHASE}}\\
	{\commentfontsize // {\em Merge partition filters to the dataset filter}}\\
	BF$_{i}$ $\leftarrow$ $\cup_{j = 1}^{|p_i|}$p-BF$_{i,j}$\;
	\Return{BF$_i$}
}

\caption{{\bf Filtering using multi-way Bloom filter}}
\label{alg:subroutines-bloomfilter-joins}
\end{algorithm}
\myparagraph{II: Multi-way Bloom filter}  
We generalize the two-way Bloom filter, so that it applies to $n$ input datasets. Consider the case where we want to perform a join operation between multiple input datasets $R_i$, where $i = 1, ..., n$: $R_1 \Join R_2 \Join ... \Join R_n$. 

Algorithm \ref{alg:subroutines-bloomfilter-joins} presents the two main steps to construct the multi-way join filter. In the first step, we create a Bloom filter BF$_i$ for each input $R_i$, where $i = 1, ..., n$ (lines 4-6), which is executed in parallel at all worker nodes that have the input datasets. In the second step, we combine the $n$ dataset filters into the join filter by simply applying the logical $AND$ operation between the dataset filters (lines 7-9). This operation adds virtually no additional overhead to build the join filter, because the logical $AND$ operation with Bloom filters is fast, even though the number of dataset filters being combined is $n$ instead of two. 

Note that an input dataset may consist of several partitions hosted on different nodes. To build the dataset filter for these partitioned inputs, we perform a simple MapReduce job that can be executed in distributed fashion: We first build the {\em partition filters} p-BF$_{i,j}$, where $j = 1, ..., |p_i|$ and $|p_i|$ is the number of partitions for input dataset $R_i$ during the Map phase, which is executed at the nodes that are hosting the partitions of each input (lines 15-21). Then, we combine the partition filters to obtain the dataset filter BF$_i$ in the Reduce phase by merging the partition filters via the logical $OR$ operation into the corresponding dataset filter BF$_i$ (lines 22-24). This process is executed for each input dataset and in parallel (see {\tt buildInputFilter()}).

\begin{figure}[t]
    \centering
    \includegraphics [scale=0.55]{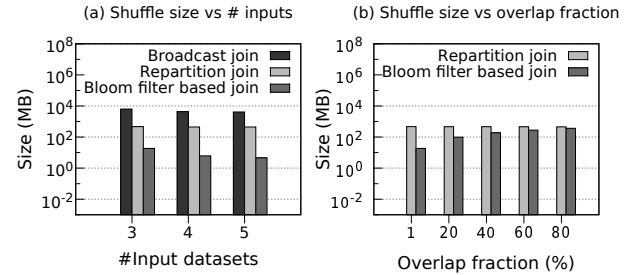} 
    \caption{Shuffled size comparison between join mechanisms: \textbf{(a)} Varying \#inputs with the overlap fraction of $1$\%; \textbf{(b)} Varying overlap fractions with three input datasets. }
\label{fig:join-mechanism-comparison}
\end{figure}

\subsubsection{Is Filtering Sufficient?}
After constructing the join filter and broadcasting it to the nodes, one straightforward approach would be to complete the join operation by performing the cross product with the data items present in the intersection. Figure~\ref{fig:join-mechanism-comparison} (a) shows the advantage of performing such a join operation with multiple input datasets based on a simulation (see \S\ref{subsec:modeling}). With the broadcast join and repartition join mechanisms, the transferred data size gradually increases with the increasing number of input datasets. However, with the Bloom filter based join approach, the transferred data size significantly reduces even when the number of datasets in the join operation increases. 

Although this filtering seems to significantly reduced transferred data among nodes, this reduction may not always be possible. Figure \ref{fig:join-mechanism-comparison} (b) shows that even with a modest overlap fraction between three input datasets (i.e., 40\%), the amount of transferred data becomes comparable with the repartition join mechanism. (In this paper, the {\em overlap fraction} is defined as the total number of data items participating in the join operation divided by the total number of data items of all inputs). Furthermore, the cross product operation will involve a significant amount of data items, potentially becoming the bottleneck. 

In \projecttitle, we first filter redundant data items as described in this section. Afterwards, we check whether the overlap fraction between
the datasets is small enough, such that we can meet the latency requirements of the user. If so, we perform the cross product of the data items participating in the join. In other words, we do not need approximation in this case (i.e., we compute the exact join result). 
If the overlap fraction is large, we continue with our approximation technique, which we describe next.

\subsection{Approximation: Cost Function}
\label{sec:cost-function}

\projecttitle supports the query budget interface for users to define a desired latency ($d_{desired}$) or a desired error bound ($err_{desired}$) as described in $\S$\ref{sec:overview}. \projecttitle ensures the join operation executed within the specified query budget by tuning the sampling parameter accordingly. In this section, we describe how \projecttitle converts the join requirements given by a user (i.e., $d_{desired}, err_{desired}$) into an optimal sampling parameter. To meet the budget, \projecttitle makes use of two types of cost functions to determine the sample size: {\em (i)} latency cost function, {\em (ii)} error bound cost function.

\myparagraph{I: Latency cost function}
In \projecttitle, we consider the latency for the join operation being dominated by two factors: 1) the time to filter and transfer participating join data items, $d_{dt}$, and 2) the time to compute the cross product, $d_{cp}$. To execute the join operation within the delay requirements of the user, we have to estimate each contributing factor.

The latency for filtering and transferring the join data items, $d_{dt}$, is measured during the filtering stage (described in $\S$\ref{sec:bloom_multiway}). We then compute the remaining allowed time to perform the join operation:
\begin{equation}
\label{eq:rem-time}
d_{rem} = d_{desired} - d_{dt}
\end{equation}
To satisfy the latency requirements, the following must hold:
\begin{equation}
\label{eq:rem-time-cross-product}
d_{cp} \leq d_{rem}
\end{equation} 

In order to estimate the latency of the cross product phase, we need to estimate how many cross products we have to perform.
Imagine that the output of the filtering stage consists of data items with $m$ distinct keys $C_{1}$, $C_{2}$  $\cdots,$ $C_{m}$. 
To fairly select data items, we perform sampling for each join key independently (explained in $\S$\ref{sec:sampling}). In other words, we will perform {\em stratified sampling}, such that each key $C_{i}$ corresponds to a stratum and has $B_{i}$ data items. Let $b_i$ represent the sample size for  $C_i$. The total number of cross products is given by:
\begin{equation}
\label{eq:cross-product-size}
CP_{total} = \sum\limits_{1}^{m} b_{i}
\end{equation}

The latency for the cross product phase would be then:
\begin{equation}
\label{eq:cross-product-delay}
d_{cp} = \beta_{compute} * CP_{total}
\end{equation}
where $\beta_{compute}$ denotes the scale factor that depends on the computation capacity of the cluster (e.g., \#cores, total memory).

\begin{figure}[t]
\centering
\includegraphics[width=0.33\textwidth]{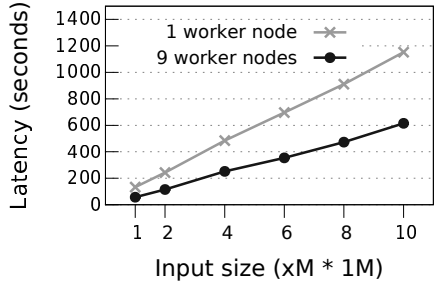}
\caption{Latency cost function: offline profiling of the compute cluster to determine $\beta_{compute}$. The plot shows latency of cross products with varying input sizes.}
\label{fig:beta-microbenchmark}
\end{figure}

We determine $\beta_{compute}$ empirically via a microbenchmark by profiling the compute cluster as an offline stage. In particular, we measure the latency to perform cross products with varying input sizes. Figure \ref{fig:beta-microbenchmark} shows that the latency is linearly correlated with the input size, which is consistent with plenty of I/O bound queries in parallel distributed settings~\cite{BlinkDB, mapreduce-cost1, mapreduce-cost2}. 
Based on this observation, we estimate the latency of the cross product phase as follows:
\begin{equation}
\label{eq:estimate-latency}
d_{cp} = \beta_{compute}*CP_{total} + \varepsilon
\end{equation} 
where $\varepsilon$ is a noise parameter. 

Given a desired latency $d_{desired}$, the sampling fraction $s = \frac{CP_{total}}{\sum\limits_{1}^{m} B_{i}}$ can be computed as: 
\begin{equation}
\label{eq:sampling-fraction}
s = (\frac{d_{rem} - \varepsilon}{\beta_{compute}})*\frac{1}{\sum\limits_{1}^{m} B_{i}} 
= (\frac{d_{desired} - d_{dt} - \varepsilon}{\beta_{compute}})*\frac{1}{\sum\limits_{1}^{m} B_{i}} 
\end{equation}

Then, the sample size $b_i$ of stratum $C_i$ can be then selected as follows:
\begin{equation}
\label{eq:samplesize-stratum-latencybound}
b_i \leq s * B_{i}
\end{equation}

According to this estimation, \projecttitle checks whether the query can be executed within the latency requirement of the user.
If not, the user is informed accordingly.

\myparagraph{II: Error bound cost function}
If the user specified a requirement for the error bound, we have to execute our sampling mechanism, such that we satisfy this requirement. Our sampling mechanism utilizes simple random sampling for each stratum (see $\S$\ref{sec:sampling}). As a result, the error $err_i$ can be computed as follows \cite{sampling-3}:
\begin{equation}\label{eq:error-stratum}
err_i = z_{\frac{\alpha}{2}} *\frac{\sigma_i}{\sqrt{b_i}}
\end{equation}
where $b_i$ represents the sample size of $C_{i}$ and $\sigma_i$ represents the standard deviation. 

Unfortunately, the standard deviation $\sigma_i$ of stratum $C_{i}$  cannot be determined without knowing the data distribution. 
To overcome this problem, we design a feedback mechanism to refine the sample size (the implementation details are in $\S$\ref{sec:implementation}): For the first execution of a query, the standard deviation of $\sigma_i$ of stratum $C_i$ is computed and stored. For all subsequent executions of the query, we utilize these stored values to calculate the optimal sample size using Equation~\ref{eq:samplesize-stratum-errbound}. 
Alternatively, one can estimate the standard deviation using a bootstrap method~\cite{BlinkDB,sampling-3}. Using this method, however, would require performing offline profiling of the data.

With the knowledge of $\sigma_i$ and solving for $b_i$ gives:
\begin{equation}\label{eq:samplesize-stratum}
b_i = (z_{\frac{\alpha}{2}}*\frac{\sigma_i}{err_i})^2
\end{equation}
With  $95$\% confidence level, we have $z_{\frac{\alpha}{2}} = 1.96$; thus, $b_i = 3.84*(\frac{\sigma_i}{err_i})^2$. $err_i$ should be less or equal to $err_{desired}$, so we have: 
\begin{equation}\label{eq:samplesize-stratum-errbound}
b_i \geq 3.84*(\frac{\sigma_i}{err_{desired}})^2
\end{equation}
Equation~\ref{eq:samplesize-stratum-errbound} allows us to calculate the optimal sample size given a desired error bound $err_{desired}$.

\myparagraph{III: Combining latency and error bound}
From Equations \ref{eq:samplesize-stratum-latencybound} and \ref{eq:samplesize-stratum-errbound}, we have a trade-off function between the latency and the error bound with confidence level of $95\%$:
\begin{equation}
\label{eq:trade-off}
d_{desired} \approx 3.84*(\frac{\sigma_i}{err_{desired}})^2*\frac{\beta_{compute}}{B_{i}}*(\sum\limits_{1}^{m} B_{i})  + d_{dt} + \varepsilon
\end{equation}

\subsection{Approximation: Sampling and Execution}
\label{sec:sampling}

In this section, we describe our sampling mechanism that executes during the cross product phase of the join operation. Executing approximation during the cross product enables \projecttitle to have highly accurate results compared to pre-join sampling. 
To preserve the statistical properties of the exact join output, we combine our technique with {\em stratified sampling}. Stratified sampling
ensures that no join key is overlooked: for each join key, we perform simple random sampling over data items independently. This method fairly selects data items from different join keys. The filtering stage ($\S$\ref{sec:bloom_multiway}) guarantees that this selection is executed only from the data items participating in the join. 

\begin{figure}[t]
\centering
\includegraphics[scale=0.33]{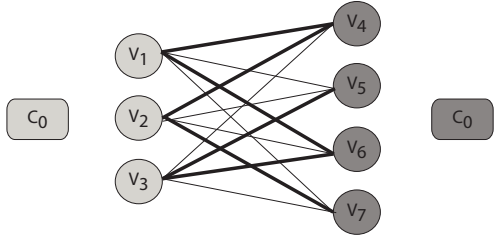}
\caption{Cross-product bipartite graph of join data items for key $C_0$. Bold lines represent sampled edges.}
\label{fig:approxjoin-sampling}
\end{figure}

For simplicity, we first describe how we perform stratified sampling during the cross product on a single node. We then describe how the sampling can be performed on multiple nodes in parallel.

\myparagraph{I: Single node stratified sampling}
Consider an inner join example of $J = R_{1} \Join R_2$ with a pair keys and values, $((k_1, v_1), (k_2, v_2))$, where $(k_1, v_1) \in R_1$ and $(k_2, v_2) \in R_2$.
This join operation produces an item $(k_1, (v_1, v_2) \in J$ if and only if $(k_1 = k_2)$. 

Consider that $R_1$ contains $(C_0, v_1)$, $(C_0, v_2)$, and $(C_0, v_3)$, and that $R_2$ contains $(C_0, v_4)$, $(C_0, v_5)$, $(C_0, v_6)$, and  $(C_0, v_7)$. The join operation based on key $C_0$ can be modeled as a complete bipartite graph (shown in Figure \ref{fig:approxjoin-sampling}). To execute stratified sampling over the join, we perform random sampling on data items having the same join key (i.e., key $C_0$). As a result, this process is equal to performing {\em edge sampling} on the complete bipartite graph. 

Sampling edges from the complete bipartite graph would require building the graph, which would correspond to computing the full cross product. To avoid this cost, we propose a mechanism to randomly select edges from the graph without building the complete graph. The function {\em sampleAndExecute()} in Algorithm~\ref{alg:sampling-graph-algo} describes the algorithm to sample edges from the complete bipartite graph. To include an edge in the sample, we randomly select one endpoint vertex from each side and then yield the edge connecting these vertices (lines 19-23). To obtain a sample of size $b_i$, we repeat this selection $b_i$ times (lines 17-18 and 24). This process is repeated for each join key $C_i$ (lines 15-24).

\myparagraph{II: Distributed stratified sampling}
The sampling mechanism can naturally be adapted to execute in a distributed setting. Algorithm~\ref{alg:sampling-graph-algo} describes how this adaptation can be achieved. In the distributed setting, the data items participating in the join are distributed to worker nodes based on the join keys using a partitioner (e.g., hash-based partitioner). A master node facilitates this distribution and directs each worker to start sampling (lines 4-5). Each worker then performs the function {\em sampleAndExecute()} in parallel to sample the join output and execute the query (lines 12-26). 

\myparagraph{III: Query execution}
After the sampling, each node executes the input query on the sample to produce a partial query result, $result_i$, and return it to the master node (lines 25-26).
The master node collects these partial results and merges them to produce a query result (lines 6-8). 
The master node also performs the error bound estimation (lines 9-10), which we describe in the following subsection (\S\ref{sec:design-error-estimation})	. 
Afterwards, the approximate query result and its error bounds are returned to the user (line 11).

\begin{algorithm}[t]

\myfontsize

%

\KwIn{
\\

$b_i$: sample size of join key $C_i$\\
$N_{1i}$ \& $N_{2i}$: set of vertices (items) in two sides of complete bipartite graph of join key $C_i$\\
$m$: number of join keys\\
$C$: set of all join keys (i.e., $\{\forall i \in \{1, ..., m\}: C_i\}$)
}

{\commentfontsize // {\em Executed sequentially at master node}}

\underline{{\bf sampleDuringJoin}}()

\Begin{
	\ForEach{$worker_i$ in $workerList$}{
		$result_i$ $\leftarrow$ $worker_i$.{\tt sampleAndExecute}();{\commentfontsize // {\em Direct workers to sample and execute the query}}\\
	}
	$result$ $\leftarrow$ $\emptyset$; {\commentfontsize // {\em Initialize empty query result}}\\ 
	\ForEach{$C_i$ in $C$}{
		$result$ $\leftarrow$ {\tt merge}($result_i$);{\commentfontsize // {\em Merge query results from workers}}\\
	}
	{\commentfontsize // {\em Estimate error for the result}}\\
	$result \pm error\_bound$ $\leftarrow$ {\tt errorEstimation}($result$)\;
	return $result \pm error\_bound$;
}

{\commentfontsize // {\em Executed in parallel at worker nodes}}

\underline{{\bf sampleAndExecute}}()

\Begin{
	
     \ForEach{$C_i$ in $C$}{
	      $sample_i$ $\leftarrow$ $\emptyset$; {\commentfontsize // {\em Sample of join key $C_i$}}\\ 
	      $count_i$ $\leftarrow$ $0$;{\commentfontsize // {\em Initialize a count to keep track \# selected items}}\\
	 
	       \While{$count_i$ $< b_i$} {
	               {\commentfontsize // {\em Select two random vertices}}\\
	               $v$ $\leftarrow$ {\tt random}($N_{1i}$)\;
	               $v'$ $\leftarrow$ {\tt random}($N_{2i}$)\;
	         
	               {\commentfontsize // {\em Add an edge between the selected vertices and update the sample}}\\
	               $sample_i$.{\tt add}($<v, v'>$)\;
	               $count_i$ $\leftarrow$ $count_i + 1$; {\commentfontsize // {\em Update counting }}\\
           }
           $result_i$ $\leftarrow$ {\tt query}($sample_i$); {\commentfontsize // {\em Execute query over sample}}\\
           return $result_i$;
       }
   }

\caption{\bf: Stratified sampling over join}
\label{alg:sampling-graph-algo}
\end{algorithm}

 \subsection{Approximation: Error Estimation}
 \label{sec:design-error-estimation}

As the final step, \projecttitle computes an error-bound for the approximate result. The approximate result is then provided to the user as $approx\-result \pm error\_bound$. 

Our sampling algorithm (i.e., the {\em sampleAndExecute()} function in Algorithm~\ref{alg:sampling-graph-algo}) described in the previous section can produce an output with duplicate edges. For such cases, we use the Central Limit Theorem to estimate the error bounds for the output. This error estimation is possible because the sampling mechanism works as a random sampling with replacement.

We can also remove the duplicate edges during the sampling process by using a hash table, and repeat the algorithm steps until we reach the desired number of data items in the sample. This approach might worsen the randomness of the sampling mechanism and could introduce bias into the sample data. In this case, we use the Horvitz-Thompson~\cite{horvitz-thompson} estimator to  remove this bias. We next explain the details of these two error estimation mechanisms.

\myparagraph{I: Error estimation using the Central Limit Theorem}
Suppose we want to compute the approximate sum of data items after the join operation. The output of the join operation contains data items with $m$ different keys $C_{1}$, $C_{2}$  $\cdots,$ $C_{m}$, each key (stratum) $C_{i}$ has $B_{i}$ data items and each such data item $j$ has an associated value  $v_{i,j}$. 
To compute the approximate sum of the join output, 
we sample $b_{i}$ items from each join key $C_{i}$ according to the parameter we computed (described in \S\ref{sec:cost-function}). 
Afterwards, we estimate the sum from this sample as $\hat{\tau} = \sum_{i=1}^{m} ( \frac{B_{i}}{b_{i}} \sum_{j=1}^{b_{i}}  v_{ij} ) \pm \epsilon$, where the error bound $\epsilon$ is defined as:
\begin{equation}\label{eq:epsilon}
\epsilon = t_{f,1-\frac{\alpha}{2}} \sqrt{\widehat{Var}(\hat{\tau})}
\end{equation}

Here, $t_{f,1-\frac{\alpha}{2}}$ is the value of the $t$-distribution (i.e., \textit{t-score}) with $f$ degrees of freedom and $\alpha = 1 - confidence level$. The degree of freedom $f$ is calculated as:
\begin{equation}\label{eq:degree_of_freedom}
f =  \sum_{i=1}^{m} b_{i} - m
\end{equation}
The estimated variance for the sum, $\widehat{Var}(\hat{\tau})$, can be expressed as:
\begin{equation}\label{eq:variance_sum}
\widehat{Var}(\hat{\tau}) = \sum\limits_{i=1}^m B_{i} * (B_{i} - b_{i}) \dfrac{r^2_{i}}{b_{i}}
\end{equation}
Here, $r^2_{i}$ is the population variance in the $i^{\textrm{th}}$ stratum. We use the statistical theories for stratified sampling~\cite{samplingBySteve} to compute the error bound.

\myparagraph{II: Error estimation using the Horvitz-Thompson estimator}
Consider the second case, where we remove the duplicate edges and resample the endpoint nodes until another edge is yielded. The bias introduced by this process can be estimated using the Horvitz-Thomson estimator.
Horvitz-Thompson is an unbiased estimator for the population sum and mean, regardless whether sampling is with or without replacement.

Let $\pi_i$ is a positive number representing the probability that data item having key $C_i$ is included in the sample under a given sampling scheme. Let $y_i$ is the sample sum of items having key $C_i$: $y_i = \sum_{j=1}^{b_{i}}  v_{ij}$. The Horvitz-Thompson estimation of the total is computed as~\cite{samplingBySteve}:
\begin{equation}\label{eq:sum-ht}
\hat{\tau_\pi}  = \sum_{i=1}^{m}( \frac{y_i}{\pi_i}) \pm \epsilon_{ht}
\end{equation}
where the error bound $\epsilon_{ht}$ is given by:
\begin{equation}\label{eq:epsilon-ht}
\epsilon_{ht} = t_{\frac{\alpha}{2}} \sqrt{\widehat{Var}(\hat{\tau_\pi})}
\end{equation}
where t has $n - 1$ degrees freedom. The estimated variance of the Horvitz-Thompson estimation is computed as:
\begin{equation}\label{eq:variance_sum-ht}
\widehat{Var}(\hat{\tau_\pi}) = \sum\limits_{i=1}^m (\frac{1 - \pi_i}{\pi_{i}^{2}})*y_i^{2} + \sum\limits_{i=1}^m \sum\limits_{j \neq i}^{} (\frac{\pi_{ij} - \pi_{i}\pi_{j}}{\pi_{i}\pi_{j}}) * \frac{y_{i}y_{j}}{\pi_{ij}}
\end{equation}
where $\pi_{ij}$ is the probability that both data items having  key $C_i$ and $C_j$ are included.

Note that the Horvitz-Thompson estimator does not depend on how many times a data item may be selected: each distinct item of the sample is used only once~\cite{samplingBySteve}.

\section{Implementation}
\label{sec:implementation}

\begin{figure}[t]
\centering
\includegraphics[scale=0.35]{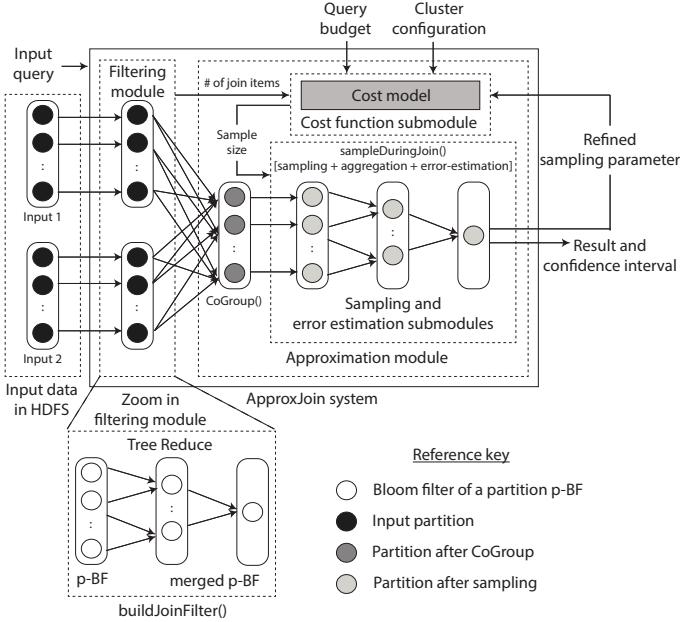}
\caption{System implementation: the figure shows distributed dataflow graph execution (on y-axis) for different stages (on x-axis)  in \projecttitle.}
\label{fig:approxjoin-implementation}
\end{figure}

In this section, we describe the implementation details of \projecttitle. 
At the high level, \projecttitle is composed of two main modules: {\em (i)} filtering and {\em (ii)} approximation. The filtering module constructs the join filter to determine the data items participating in the join. These data items are fed to the approximation module to perform the join query within the query budget  specified by the user. 

We implemented our design by modifying Apache Spark~\cite{spark}. Spark uses Resilient Distributed Datasets (RDDs)~\cite{spark-nsdi-2012} for scalable and fault-tolerant distributed data-parallel computing. An RDD is an immutable collection of objects distributed across a set of machines. To support existing programs, we provide a simple programming interface that is also based on the RDDs. In other words, all operations in \projecttitle, including filtering and approximation, are transparent to the user. To this end, we have implemented a PairRDD for {\em approxjoin()} function to perform the join query within the query budget over inputs in the form of RDDs. Figure~\ref{fig:approxjoin-implementation} shows in detail the directed acyclic graph (DAG) execution of \projecttitle. 
 
\begin{figure*}[t]
\centering
\includegraphics [width=1\textwidth]{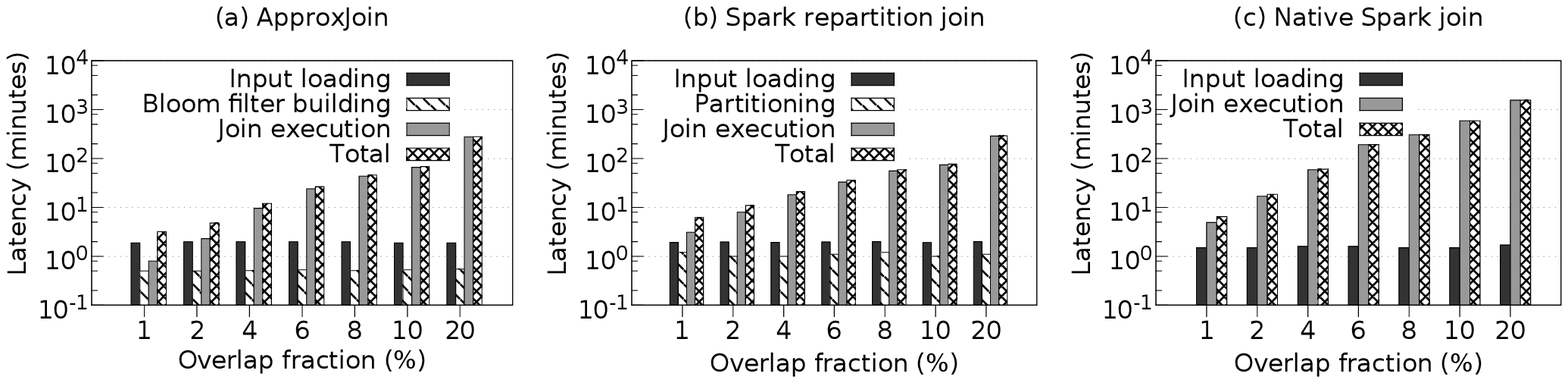}
\caption{Benefits of filtering in two-way joins.  We show the total latency and the breakdown latency of \textbf{(a)} \projecttitle, \textbf{(b)} Spark repartition join, and \textbf{(c)} native Spark join.}
\label{fig:micro-benchmarks-1}
\end{figure*}

\myparagraph{I: Filtering module} The join Bloom filter module implements the filtering stage described in \S\ref{sec:bloom_multiway} to eliminate the non-participating data items. A straightforward way to implement {\em buildJoinFilter()} in Algorithm~\ref{alg:subroutines-bloomfilter-joins} is to build Bloom filters for all partitions (p-BFs) of each input and merge them in the driver of Spark in the Reduce phase. However, in this approach,  the driver quickly becomes a bottleneck when there are multiple data partitions located on many workers in the cluster.  To solve this problem, we leverage the {\em treeReduce} scheme~\cite{incoop, slider}.  In this model, we combine the Bloom filters in a hierarchical fashion, where the reducers are arranged in a tree with the root performing the final merge (Figure~\ref{fig:approxjoin-implementation}). If the number of workers increases (i.e., \projecttitle deployed in a bigger cluster), more layers are added to the tree to ensure that the load on the driver remains unchanged. After building the join filter, \projecttitle broadcasts it to determine participating join items in all inputs and feed them to the approximation module.

The approximation module consists of three submodules including  the cost function, sampling and error estimation.
The cost function submodule implements the mechanism in $\S$\ref{sec:cost-function} to determine the sampling parameter according to the requirements in the query budget. The sampling submodule performs the proposed sampling mechanism (described in $\S$\ref{sec:sampling}) and executes the join query over the filtered data items with the sampling parameter.  The error estimation submodule computes the error-bound (i.e., confidence interval) for the query result from the sampling module (described in $\S$\ref{sec:design-error-estimation}). This error estimation submodule also performs fine-tuning of the sample size used by the sampling submodule to meet the accuracy requirement in subsequent runs.

\myparagraph{II: Approximation: Cost function submodule}  The cost function submodule converts the query budget requirements provided by the user into the sampling parameter used in the sampling submodule. We implemented a simple cost function by building a model to convert the desired latency into the sampling parameter. To build the model, we perform offline profiling of the compute cluster. This model empirically establishes the relationship between the input size and the latency of cross product phase by computing the $\beta_{compute}$ parameter from the microbenchmarks. Afterwards, we utilize Equation~\ref{eq:samplesize-stratum-latencybound} to compute the sample sizes.

\myparagraph{III: Approximation: Sampling submodule} After receiving the intersection of the inputs from the filtering module and the sampling parameter from the cost function submodule, the sampling submodule performs the sampling during the join as described in \S\ref{sec:sampling}. We implemented the proposed sampling mechanism in this submodule by creating a new Spark PairRDD function {\em sampleDuringJoin()} that executes stratified sampling during the join.

The original {\em join()} function in Spark uses two operations: 1)  {\em cogroup()} shuffles the data in the cluster, and 2)  {\em cross-product} performs the final phase in join. In our {\em approxjoin()} function, we replace the second operation with {\em sampleDuringJoin()} that implements our mechanism described in \S\ref{sec:sampling} and Algorithm \ref{alg:sampling-graph-algo}. Note that the data shuffled by the {\em cogroup()} function is the output of the filtering stage. As a result, the amount of shuffled data can be significantly reduced if the overlap fraction between datasets is small. Another thing to note is that {\em sampleDuringJoin()} also performs the query execution as described in Algorithm~\ref{alg:sampling-graph-algo}.

\myparagraph{IV: Approximation: Error estimation submodule} After the query execution is performed in {\em sampleDuringJoin()}, the error estimation submodule implements the function {\em errorEstimation()} to compute the error bounds of the query result. The submodule also activates a feedback mechanism to re-tune the sample sizes in the sampling submodule to achieve the specified accuracy target as described in \S\ref{sec:cost-function}. We use the Apache Common Math library~\cite{math-apache} to implement the error estimation mechanism described in \S\ref{sec:design-error-estimation}.

\section{Evaluation: Microbenchmarks}
\label{sec:evaluation}

\begin{figure*}[t]
    \centering
    \includegraphics [width= 1\textwidth]{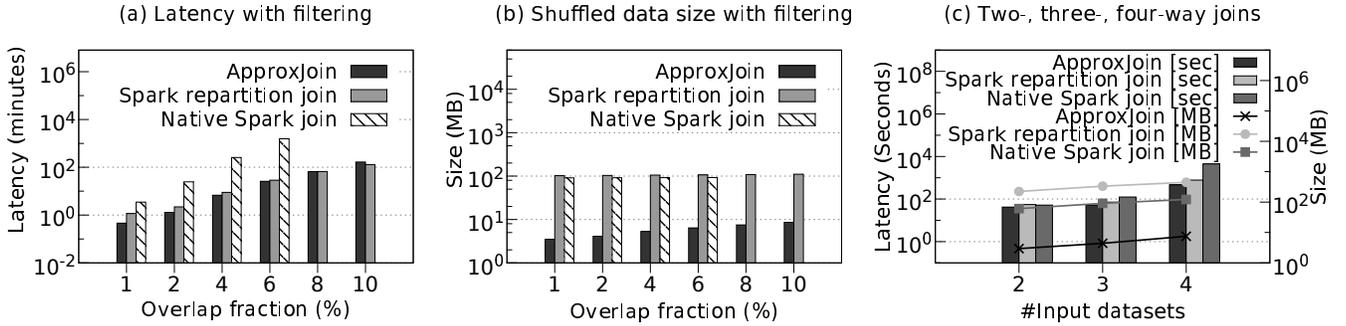} 
    \caption{Benefits of filtering in multi-way joins, with different overlap fractions and different numbers of input datasets.}
          \vspace{-2mm}
\label{fig:micro-benchmarks-3}
\end{figure*}

In this section, we present the evaluation results of \projecttitle based on microbenchmarks and the TPC-H benchmark. In the next section, we will report evaluation based on real-world case studies.

\subsection{Experimental Setup}
\label{subsec:evaluation-setup}

\myparagraph{Cluster setup} 
Our cluster consists of $10$ nodes, each equipped with two Intel Xeon E5405 quad-core CPUs,  $8$GB memory and a SATA-2 hard disk, running Ubuntu $14.04.1$.

\myparagraph{Synthetic datasets} We analyze the performance of \projecttitle using synthetic datasets following Poisson distributions with $\lambda$ in the range of $[10,10000]$.
The number of distinct join keys is set to be proportional to the number of workers.

\myparagraph{Metrics} We evaluate \projecttitle using three metrics: latency, shuffled data size, and accuracy loss. Specifically, the latency is defined as the total time consumed to process the join operation; 
the shuffled data size is defined as the total size of the data shuffled across nodes during the join operation;
the accuracy loss is defined as $(approx - exact) / exact$, where $approx$ and $exact$ denote the results from the executions with and without sampling, respectively. 

\subsection{Benefits of Filtering}
\label{subsec:eval-overlap}

The join operation in \projecttitle consists of two main stages: {\em (i)} filtering stage for reducing shuffled data size,  and {\em (ii)} sampling stage for approximate computing. In this section, we activate only the filtering stage (without the sampling stage) in \projecttitle, and evaluate the benefits of the filtering stage.

\myparagraph{I: Two-way joins} First, we report the evaluation results with two-way joins.
Figure~\ref{fig:micro-benchmarks-1}(a)(b)(c) show the latency breakdowns of \projecttitle, Spark repartition join, and native Spark join, respectively. Unsurprisingly, the results show that building bloom filters in \projecttitle is quite efficient (only around $42$ seconds) compared with the cross-product-based join execution (around $43\times$ longer than building bloom filters, for example, when the overlap fraction is $6$\%).  The results also show that the cross-product-based join execution is fairly expensive across all three systems.

When the overlap fraction is less than $4$\%, \projecttitle achieves $2 \times$ and $3 \times$ shorter latencies than Spark repartition join and native Spark join, respectively. 
However, with the increase of the overlap fraction, there is an increasingly large amount of data that has to be shuffled and the expensive cross-product operation cannot be eliminated in the filtering stage; therefore, the benefit of the filtering stage in \projecttitle gets smaller.
For example, when the overlap fraction is $10$\%, \projecttitle speeds up only $1.06 \times$ and $8.2 \times$ compared with Spark repartition join and Spark native join, respectively.  When the overlap fraction increases to 20\%, \projecttitle's latency does not improve (or may even perform worse) compared with the Spark repartition join.  At this point, we need to activate the sampling stage of \projecttitle to reduce the latency of the join operation, which we will evaluate in \S\ref{sec:eval-sampling}.

\begin{figure*}[t]
    \centering
    \includegraphics [width=1\textwidth]{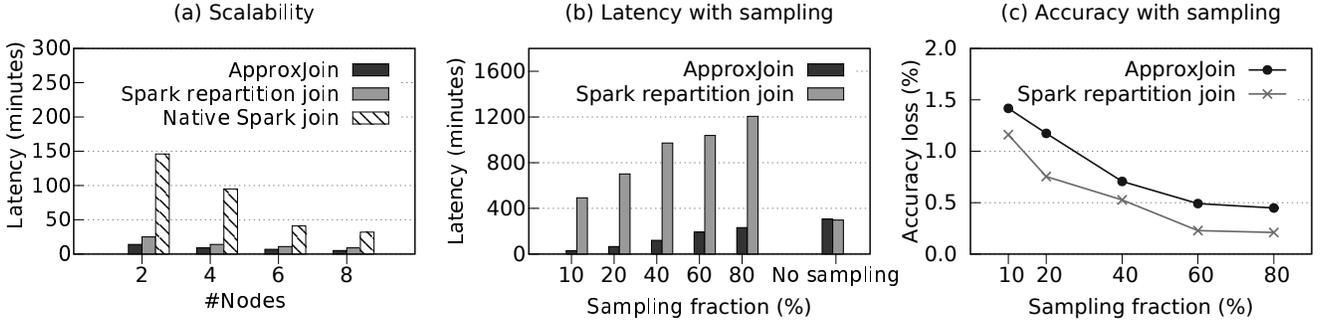} 
    \caption{Comparison between \projecttitle and Spark join systems in terms of \textbf{(a)} scalability, \textbf{(b)} latency with sampling, and \textbf{(c)} accuracy loss with sampling.}
     
          \vspace{-3mm}
\label{fig:micro-benchmarks-2}
\end{figure*}
\begin{figure}[t]
    \centering
    \includegraphics [width=0.48\textwidth]{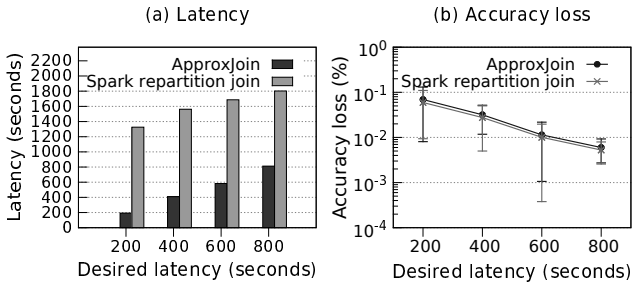} 
    \caption{Effectiveness of the cost function.} 
\label{fig:costfunction-evaluation}
\end{figure}

\myparagraph{II: Multi-way joins}  Next, we present the evaluation results with multi-way joins. Specifically, we first conduct the experiment with three-way joins whereby we create three synthetic datasets with the same aforementioned Poisson distribution. 

We measure the latency and the shuffled data size during the join operations in \projecttitle, Spark repartition join and native Spark join, with varying overlap fractions. Figure~\ref{fig:micro-benchmarks-3}(a) shows that,
with the overlap fraction of $1$\%, \projecttitle is $2.6\times$ and $8\times$ faster than Spark repartition join and native Spark join, respectively. However, with the overlap fraction larger than $8$\%, \projecttitle does not achieve much latency gain (or may even perform worse) compared with Spark repartition join.  This is because, similar to the two-way joins, the increase of the overlap fraction prohibitively leads to a larger amount of data that needs to be shuffled and cross-producted.  Note also that, we do not have the evaluation results for native Spark join with the overlap fractions of $8\%$ and $10\%$, simply because that system runs out of memory. 
In addition, Figure~\ref{fig:micro-benchmarks-3}(b) shows that \projecttitle significantly reduces the shuffled data size. For example, with the overlap fraction of $6\%$, \projecttitle reduces  the shuffled data size by $16.68\times$ and $14.5\times$ compared with Spark repartition join and native Spark join, respectively.

Next, we conduct experiments with two-way, three-way and four-way joins.  In two-way joins, we use two synthetic datasets A and B that have an overlap fraction of $1$\%; in three-way joins, the three synthetic datasets A, B, and C have an overlap fraction of $0.33$\%, and the overlap fraction between any two of them is also $0.33$\%; in four-way joins, the four synthetic datasets have an overlap fraction of $0.25$\%, and the overlap fraction between any two of these datasets is also $0.25$\%.

Figure~\ref{fig:micro-benchmarks-3}(c) presents the latency and the shuffled data size during the join operation with different numbers of input datasets. With two-way joins, \projecttitle speeds up by $2.2\times$ and $6.1\times$, and reduces the shuffled data size by $45\times$ and $12\times$, compared with Spark repartition join and native Spark join, respectively. In addition, with three-way and four-way joins, \projecttitle achieves even larger  performance gain.  This is because, with the increase of the number of input datasets, the number of non-join data items also increases; therefore, \projecttitle gains more benefits from the filtering stage.

\myparagraph{III: Scalability}
Finally, we keep the overlap fraction of $1\%$ and evaluate the scalability of \projecttitle with different numbers of compute nodes. Figure~\ref{fig:micro-benchmarks-2}(a) shows that \projecttitle achieves a lower latency than Spark repartition join and native Spark join.  With two nodes, \projecttitle achieves a speedup of $1.8\times$ and $10\times$  over Spark repartition join and native Spark join, respectively. Meanwhile, with $8$ nodes, \projecttitle achieves a speedup of $1.7\times$ and $6\times$ over Spark repartition join and native Spark join.

\subsection{Benefits of Sampling}
\label{sec:eval-sampling}
\begin{figure*}[t]
    \centering
    \includegraphics [width= 1\textwidth]{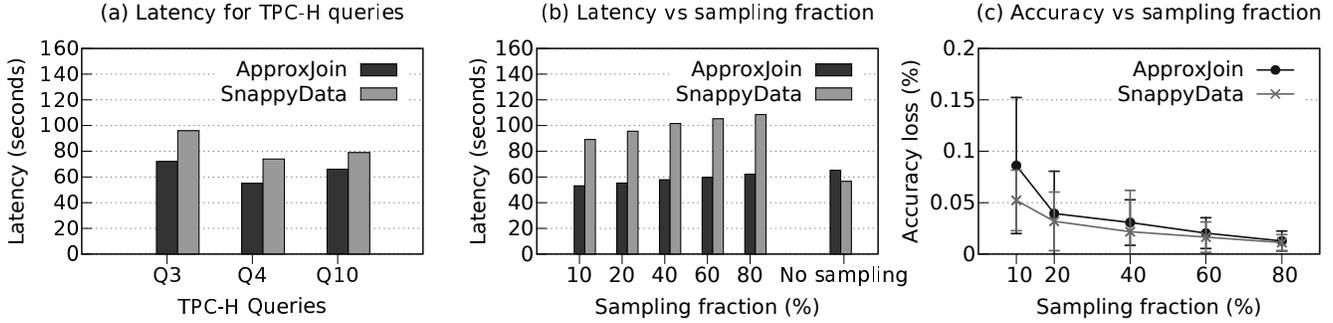} 
    
      \caption{Comparison between \projecttitle and the state-of-the-art SnappyData system in terms of \textbf{(a)} latency with different TPC-H queries, \textbf{(b)} latency with different sampling fractions, and \textbf{(c)} accuracy with different sampling fractions.}
\label{fig:micro-benchmarks-4}
\end{figure*}

As shown in previous experiments, \projecttitle does not gain much latency benefit from the filtering stage when the overlap fraction is large. To reduce the latency of the join operation in this case, we activate the second stage of \projecttitle, i.e., the sampling stage.  

For a fair comparison, we re-purpose Spark's built-in sampling algorithm (i.e., stratified sampling via {\tt sampleByKey}) to build a ``sampling over join'' mechanism for the Spark repartition join system.  Specifically, we perform the stratified sampling over the join results after the join operation has finished in the Spark repartition join system.
We then evaluate the performance of \projecttitle, and compare it with this \emph{extended} Spark repartition join system.

\myparagraph{I: Latency} We measure the latency of \projecttitle and the extended Spark repartition join with varying sampling fractions.
Figure~\ref{fig:micro-benchmarks-2}(b) shows that the Spark repartition join system scales poorly with a significantly higher latency as it could perform stratified sampling only after finishing the join operation.  Even if we were to enable the Spark repartition join system to perform stratified sampling over the input datasets and then perform the join operation over these samples, this would come with a significant accuracy loss.

\myparagraph{II: Accuracy}  Next, we measure the accuracy of \projecttitle and the extended Spark repartition join. Figure~\ref{fig:micro-benchmarks-2}(c) shows that the accuracy losses in both systems decrease with the increase of sampling fractions, although \projecttitle's accuracy is slightly worse than the Spark repartition join system.  Note however that, as shown in Figure~\ref{fig:micro-benchmarks-2}(b), \projecttitle achieves an order of magnitude speedup compared with the Spark repartition join system since \projecttitle performs sampling during the join operation.

\subsection{Effectiveness of the Cost Function}
\projecttitle provides users with a query budget interface, and uses a cost function to convert the query budget into a sample size (see $\S$\ref{sec:cost-function}).
In this experiment, a user sends \projecttitle  a join query along with a latency budget (i.e., the desired latency the user wants to achieve). \projecttitle uses the cost function, whose parameter is set according to the microbenchmarks ($\beta = 4.16*10^{-9}$ in our cluster), to convert the desired latency to the sample size. We measure the latency of \projecttitle and the extended Spark repartition join in performing the join operations with the identified sample size.  Figure~\ref{fig:costfunction-evaluation}(a) shows that \projecttitle can rely on the cost function to achieve the desired latency quite well (with the maximum error being less than 12 seconds). Note also that, the Spark repartition join incurs a much higher latency than \projecttitle since it performs the sampling after the join operation has finished. In addition, Figure~\ref{fig:costfunction-evaluation}(b) shows that  \projecttitle can achieve a very similar accuracy to the Spark repartition join system.

\subsection{Comparison with SnappyData using TPC-H}
\label{subsec:tpc-h}

In this section, we evaluate \projecttitle using TPC-H benchmark. TPC-H benchmark consists of $22$ queries, and has been widely used to evaluate various database systems.  We compare \projecttitle with the state-of-the-art related system --- SnappyData~\cite{snappydata}.

SnappyData is a hybrid distributed data analytics framework which supports a unified programming model for transactions,  OLAP and data stream analytics. It integrates GemFine, an in-memory transactional store, with Apache Spark. SnappyData inherits approximate computing techniques from BlinkDB~\cite{BlinkDB} (off-line sampling techniques) and the data synopses to provide  interactive analytics. SnappyData does not support sampling over joins.

In particular, we compare \projecttitle with SnappyData  using the TPC-H queries $Q3$, $Q4$ and $Q10$ which contain join operations. To make a fair comparison, we only keep the join operations and remove other operations in these queries. We run the benchmark with a scale factor of $10\times$, i.e., $10$GB datasets.

First, we use the TPC-H benchmark to analyze the performance of \projecttitle with the filtering stage but without the sampling stage. Figure~\ref{fig:micro-benchmarks-4}(a) shows the end-to-end latencies of \projecttitle and SnappyData in processing the three queries. \projecttitle is $1.34\times$ faster than SnappyData in processing $Q4$ which contains only one join operation. In addition, for the query $Q3$ which consists of two join operations, \projecttitle achieves a  $1.3\times$ speedup than SnappyData. Meanwhile,  \projecttitle speeds up by $1.2\times$ compared with SnappyData for the query $Q10$. 

Next, we evaluate \projecttitle with both filtering and sampling stages activated. In this experiment, we perform a query to answer the question {\em ``what is the total amount of money the customers had before ordering?''}. To process this query, we need to join the two tables $CUSTOMER$ and $ORDERS$ in the TPC-H benchmark, and then sum up the two fields $o\_totlaprice$   and $c\_acctbal$.

Since SnappyData does not support sampling over the join operation, in this experiment it first executes the join operation between the two tables $CUSTOMER$ and $ORDERS$, then performs the sampling over the join output, and finally calculates the sum of the two fields $o\_totalprice$  and $c\_acctbal$.
Figure~\ref{fig:micro-benchmarks-4}(b) presents the latencies of \projecttitle and  SnappyData in processing the aforementioned query with varying sampling fractions. 
SnappyData has a significantly higher latency than \projecttitle, simply because it performs sampling only after the join operation finishes. 
For example, with a sampling fraction of $60\%$, SnappyData achieves a $1.77\times$ higher latency  than \projecttitle, even though it is faster when both systems do not perform  sampling (i.e., sampling fraction is $100\%$).  Note however that, sampling is inherently needed when one  handles  joins with large-scale inputs that require a significant number of cross-product operations.
Figure~\ref{fig:micro-benchmarks-4}(c) shows the accuracy losses of \projecttitle and SnappyData. \projecttitle achieves an accuracy level similar to SnappyData. For example, with a sampling fraction of $60\%$, \projecttitle achieves an accuracy loss of $0.021\%$, while SnappyData achieves an accuracy loss of $0.016\%$.

\section{Evaluation: Real-world Datasets}
\label{sec:case-studies}

\begin{figure*}[t]
\centering
\includegraphics [width=1\textwidth]{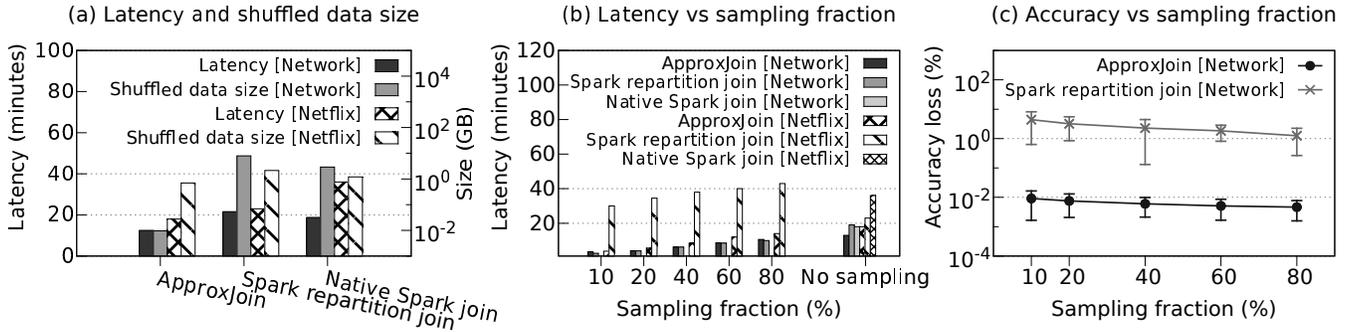}
\caption{Comparison between \projecttitle, Spark repartition join, and native Spark join based on two real-world datasets: (1) Network traffic monitoring dataset (denoted as [Network]), and (2) Netflix Prize dataset (denoted as [Netflix]).}
\label{fig:case-studies}
\end{figure*}

We evaluate \projecttitle based on two real-world datasets: (a) network traffic monitoring dataset and (b) Netflix Prize dataset.

\subsection{Network Traffic Monitoring Dataset}

\label{subsec:eval-case-study-network}

\myparagraph{Dataset} 
We use the CAIDA network traces~\cite{caida2015} which were collected  on the Internet backbone links in Chicago in 2015. 
In total, this dataset contains $115,472,322$ TCP flows, $67,098,852$ UDP flows, and $2,801,002$ ICMP flows.
Here, a flow denotes a two-tuple network flow that has the same source and destination IP addresses.

\myparagraph{Query} We use \projecttitle to process the query: {\em What is the total size of the flows that appeared in all TCP, UDP and ICMP traffic?} To answer this query, we need to perform a join operation across TCP, UDP and ICMP flows.

\myparagraph{Results} Figure~\ref{fig:case-studies}(a) first shows the latency comparison between \projecttitle (with filtering but without sampling), Spark repartition join, and native Spark join. \projecttitle achieves a latency  $1.72\times$ and $1.57\times$ lower than Spark repartition join and native Spark join, respectively. Interestingly, native Spark join achieves a lower latency than Spark repartition join. This is because the dataset is distributed quite uniformly across worker nodes in terms of the join-participating flow items, i.e., there is little data skew. Figure~\ref{fig:case-studies}(a) also shows that \projecttitle significantly reduces the shuffled data size by a factor of $300\times$ compared with the two Spark join systems.

Next, different from the experiments in \S\ref{sec:evaluation}, we extend Spark repartition join by enabling it to sample the dataset before the actual join operation.  This leads to the lowest latency it could achieve.  Figure~\ref{fig:case-studies}(b) shows that \projecttitle achieves a similar latency even to this extended Spark repartition join.  In addition, Figure~\ref{fig:case-studies}(c)  shows the accuracy loss comparison between \projecttitle and Spark repartition join with different sampling fractions. As the sampling fraction increases, the accuracy losses of \projecttitle and Spark repartition join decrease, but not linearly. \projecttitle produces around $42\times$ more accurate query results  than the Spark repartition join system with the same sampling fraction. 

\subsection{Netflix Prize Dataset}
\label{subsec:eval-case-study-netflix}

\myparagraph{Dataset} We also evaluate \projecttitle based on the Netflix Prize dataset which includes around $100M$ ratings of $17,770$ movies by $480,189$ users. Specifically, this dataset contains $17,770$ files, one per movie, in the $training\_set$ folder. The first line of each such file contains $MovieID$, and each subsequent line in the file corresponds to a rating from a user and the date, in the form of $\langle UserID, Rating, Date \rangle$.  There is another file
$qualifying.txt$ which contains lines indicating $MovieID$, $UserIDs$ and the rating $Dates$.

\myparagraph{Query} We perform the join operation between the dataset in $training\_set$ and the dataset in $qualifying.txt$ to evaluate \projecttitle in terms of latency. Note that, we cannot find a meaningful aggregation query for this dataset; therefore, we focus on only the latency but not the accuracy of the join operation.

\myparagraph{Results} Figure~\ref{fig:case-studies}(a) shows the latency and the shuffled data size of \projecttitle (with filtering but without sampling), Spark repartition join, and native Spark join. \projecttitle is $1.27\times$ and $2\times$ faster than Spark repartition join and native Spark join, respectively. The result in Figure~\ref{fig:case-studies}(a) also shows that \projecttitle reduces the shuffled data size by $3\times$ and $1.7\times$ compared with Spark repartition join and native Spark join, respectively.
In addition, Figure~\ref{fig:case-studies}(b) presents the latency comparison between these systems with different sampling fractions. For example, with the sampling fraction of $10$\%, \projecttitle is $6\times$ and $9\times$ faster than Spark  repartition join and native Spark join, respectively.
Even without sampling (i.e., sampling fraction is $100\%$), \projecttitle is still $1.3\times$ and $2\times$ faster than Spark repartition join and native Spark join, respectively.

\section{Related Work}
\label{sec:related}

Over the last decade, approximate computing has been applied in various domains
such as programming languages~\cite{green,enerj}, hardware
design~\cite{approxhardware1}, and distributed systems~\cite{hop,online-aggregation,online-aggregation-mapreduce}. Our techniques mainly target the databases research community~\cite{aqua, BlinkDB, quickr-sigmod, daq, cs2, stratified-sampling-joins-1, stratified-sampling-joins-2, incapprox-www-2016, streamapprox-middleware17, privapprox-atc17, approxiot-icdcs-2018, privapprox-tech-report, streamapprox-tech-report}. 
In particular, various approximation techniques have been proposed to make trade-offs between
required resources and output quality, including
sampling~\cite{stratified-sampling, sampling-2}, sketches~\cite{sketching}, and
online aggregation~\cite{online-aggregation, online-aggregation-mapreduce}. 
Chaudhari et al. provide a sampling over join mechanism by taking a sample of an input and considering all statistical characteristics and indices of other inputs~\cite{samplingoverjoin}.  AQUA~\cite{aqua} system makes use of simple random sampling to take a sample of joins of inputs that have primary key-foreign key relations. BlinkDB~\cite{BlinkDB} proposes an approximate distributed query processing engine that uses stratified sampling~\cite{stratified-sampling} to support ad-hoc queries with error and response time constraints. SnappyData~\cite{snappydata} and SparkSQL~\cite{spark-sql} adopt the approximation techniques from BlinkDB to support approximate queries. 
Quickr~\cite{quickr-sigmod} deploys distributed sampling operators to reduce execution costs of parallel, ad-hoc queries that may contain multiple join operations. In particular, Quickr first injects sampling operators into the query plan and searches for an optimal query plan among sampled query plans to execute input queries.

Unfortunately, all of these systems require a priori knowledge of the inputs. For example, AQUA~\cite{aqua} requires join inputs to have primary key-foreign key relations. For another example, the sampling over join mechanism~\cite{samplingoverjoin} needs the statistical characteristics and indices of inputs. Finally, BlinkDB~\cite{BlinkDB} utilizes most frequently used column sets to perform off-line stratified sampling over them. Afterwards, the samples are cached, such that queries can be served by selecting the relevant samples and executing the queries over them. While useful in many applications, BlinkDB and these other systems cannot process queries over new inputs, where queries or inputs are typically not known in advance. 

Ripple Join~\cite{ripple-join} implements online aggregation for joins. Ripple Join repeatedly takes a sample from each input. For every item selected, it is joined with all items selected in other inputs so far. Recently, Wander Join~\cite{wander-join} improves over Ripple Join by performing random walks over the join data graph of a multi-way join. However, their approach crucially depends on the availability of indices, which are not readily available in ``big data" systems like Apache Spark. In addition, the current Wander Join implementation is single-threaded, and parallelizing the walk plan optimization procedure is non-trivial.
In this work, we proposed a simple but efficient sampling mechanism over joins which works not only on a single node but also in a distributed setting.

Recently, an approximate query processing (AQP) formulation~\cite{AQP} has been proposed to provide low-error approximate results without any preprocessing or a priori knowledge of inputs.  The formulation based on probability theory allows to reuse results of past queries to improve the performance of future query processing. However, the current version of AQP formulation does not support joins.

\section{Conclusion}
\label{sec:conclusion}

The keynote speakers at SIGMOD 2017~\cite{Kraska-keynote, AQP-Chaudhuri1, Mozafari-keynote} highlighted the challenges and opportunities in approximate query processing. In a follow up succinct blog post~\cite{AQP-Chaudhuri2}, Chaudhuri explains the reasons why, in spite of decades of technical results,  the problem of approximate joins is hard even for a simple join query with group-by and aggregation. In this work, we strive to address the challenges associated in performing approximate joins for distributed data analytics systems.  We achieve this by performing sampling during the join operation to achieve low latency as well as high accuracy. In particular, we employed a sketching technique (Bloom filters) to reduce the size of the shuffled data during the joins, and also proposed a stratified sampling mechanism that executes during the join in a distributed setting. We implemented our techniques in a system called \projecttitle using Apache Spark and evaluated its effectiveness using a series of microbenchmarks and real-world case studies. Our evaluation shows that \projecttitle significantly reduces query response times as well as data shuffled through the network without losing accuracy of the query results compared with the state-of-the-art systems.

\myparagraph{Supplementary material} The appendix contains analysis of \projecttitle covering both communication and computation complexities (Appendix~\ref{sec:approxjoin-analysis}). In addition, we also discuss three alternative design choices for Bloom filters  (Appendix~\ref{sec:discussion}).

\balance
\newpage
\bibliographystyle{abbrv}
\bibliography{main}

\newpage
\appendix
\section*{Appendix}
\section{Analysis of ApproxJoin}
\label{sec:approxjoin-analysis}

\begin{table}[]
	\centering
	\begin{tabular}{|l|l|}
		\hline
		{\bf Symbol}            & {\bf Meaning}                                \\
		\hline
		R$_i$             & A join input \\
		$n$                 & The number of join inputs \\
		$m$                & The number of join keys \\
		$k$                 & The number of worker nodes\\
		$h$                 & The number of hash functions in bloom filters\\
		BF$_i$            & Bloom filter of input $R_i$            \\
		p-BF$_{i,j}$ & Bloom filter of partition $p_{i,j}$ in $R_i$   \\
		join filter       & The global Bloom filter for all inputs \\
		$d_{desired}$ & Desired latency \\
		$err_{desired}$ & Desired error \\
		$C_{i}$              & A join key (a stratum) \\
		$B_{i}$              & The total number of data items having join key $C_{i}$ \\
		$b_i$                 & $b_{i}$ sample size of join key $C_{i}$ \\
		$d_{dt}$            & Data transfer delay \\
		$d_{cp}$           & Cross-product computing delay \\
		$DT_{total}$     & Total data transfered size \\
		$CP_{total}$     & Total cross-product size \\
		$\beta_{compute}$  & Parameter for execution environment\\
		$err_i$               & The error bound of a join key $C_{i}$ \\
		$\varepsilon$    & Noise parameter of execution environment \\
		$|p_i|$                 & The number of partitions of input $R_i$ \\
		\hline 
	\end{tabular}
	\vspace{0.5em}
	\caption{Symbols and terms used in this paper}
	\label{my-label}
\end{table}


In this section, we first analyze the communication complexity of \projecttitle. Thereafter, we provide computational complexity analysis for the proposed stratified sampling over joins in the sampling stage of \projecttitle.

\subsection{Communication Complexity}
\label{subsec:modeling}

For the communication complexity, we analyze the performance gain of \projecttitle in terms of shuffled data size during the filter stage with various setting of Bloom filters using a model-based analysis. We compare the gains of \projecttitle with the broadcast and repartition join mechanisms. Based on our analysis, we also show how to select input parameters for Bloom filter to achieve an optimal trade-off between reducing the shuffled data volume and the desired false positive value in the Bloom filters.

Suppose we want to  execute a multi-way join operation on attribute $A$ for $n$ input datasets $R_1 \Join R_2 \Join ... \Join R_n$, where $R_i (i = 1, ..., n)$ is an input dataset. For simplicity, we assume that $|R_1| < |R_2| < ... < |R_n|$. The number of nodes in our experimental cluster is $k$ and $k > 1$.

\myparagraph{I: Broadcast join} In broadcast join, we broadcast all smaller datasets to all nodes that contain the largest dataset. The total shuffled data volume is bounded by:
\begin{equation}
\label{eq:broadcast-shuffle-data}
S_{bc} = \langle |R_1| + |R_2| + ... + |R_{n-1}|  \rangle * (k -1)
\end{equation}
When we add one more node to the cluster, the relative increase in the shuffled data volume in broadcast join will be:
\begin{equation}
\frac{\theta_F}{\theta_k} =  |R_1| + |R_2| + ... + |R_{n-1}| 
\end{equation}
When we add one more dataset $R_{n+1}$ to the join operation, the relative increase in the shuffled data volume will be:
\begin{equation}
\frac{\theta_F}{\theta_n} =  |R_n| * (k -1)
\end{equation}

\myparagraph{II: Repartition join} In repartition join, we shuffle data items of datasets across the cluster to make sure that each node in the cluster will keep at least a chunk/partition of each dataset.  Therefore, the shuffled data volume in repartition join is computed as follows:
\begin{equation}
\label{eq:repartition-shuffle-data}
S_{re} = \langle |R_1| + |R_2| + ... + |R_{n-1}| +|R_n|  \rangle * \frac{k -1}{k}
\end{equation}
When we add one more node to the cluster, the relative increase in the shuffled data volume in repartition join will be:
\begin{equation}
\frac{\theta_F}{\theta_k} =  \langle |R_1| + |R_2| + ... + |R_{n-1}| + |R_n| \rangle * \frac{1}{k*(k + 1)}
\end{equation}
When we add one more dataset $R_{n+1}$ to the join operation, the relative increase in the shuffled data volume will be:
\begin{equation}
\frac{\theta_F}{\theta_n} =  |R_{n+1}| * \frac{k -1}{k}
\end{equation}

\myparagraph{III: \projecttitle} Algorithm \ref{alg:subroutines-bloomfilter-joins} describes our proposed filtering using a Bloom-filter for multi-way joins. The algorithm builds a Bloom filter BF$_i$ for each dataset $R_i$ using the function {\em buildInputFilter}. In the Map phase, the function builds Bloom filters for all partitions of the input dataset $R_i$.  In the Reduce phase, all the partitioned Bloom filters are merged to build the Bloom filter  BF$_i$ for the input dataset $R_i$. Since we fix the size for all Bloom filters, the volume of the shuffled data for building Bloom filters for all inputs is computed as $ |BF|*(k -1)*n$, where $|BF|$ is the size of each Bloom filter. Thereafter, the Bloom filters of all inputs are combined to build a join Bloom filter with size $|BF|$ for all join input using the function {\em buildJoinFilter}.

Next, the algorithm broadcasts the join Bloom filter to all nodes to filter out all data items that do not participate in the join operation. The shuffled data size of the broadcast step is calculated as $|BF|*( k - 1)$. The volume of shuffled data for the filtering step is computed as $\langle |r_1| + |r_2| + ... +|r_n| \rangle * \frac{k - 1}{k}$; where $|r_i|$ is the size of data items participating in the join operation of input $R_i$ .

In summary, the total volume of shuffled data in the proposed filtering mechanism is calculated as follows:
\begin{equation}
\label{eq:bloomfilter-shuffle-data}
\begin{split}
S_{bf} =  &\quad   |BF|*(k - 1) *(n + 1) +  \\ 
              &\quad  + \langle |r_1| + |r_2| + ...  +|r_n|  \rangle * \frac{k -1}{k} 
  \end{split}
\end{equation}
When we add one more node to the cluster, the relative increase in the shuffled data volume in \projecttitle will be:
\begin{equation}
\begin{split}
\frac{\theta_F}{\theta_k} =  &\quad   |BF| *(n + 1) + \\
                                          &\quad  + \langle |r_1| + |r_2| + ... + |r_{n-1}| + |r_n| \rangle * \frac{1}{k*(k + 1)}
  \end{split}
\end{equation}
When we add one more dataset $R_{n+1}$ to the join operation, the relative increase of the shuffled data volume is computed as:
\begin{equation}
\frac{\theta_F}{\theta_n} =  |BF| *(k - 1) +  |r_{n+1}| * \frac{k -1}{k}
\end{equation}
 
 Note that for Bloom filters, false positives are possible, but false negatives are not. There is a trade-off between the size of a bit vector $|BF|$ and the probability of a false positive. A larger $|BF|$  has fewer false positives but consumes more memory, whereas a smaller  $|BF|$  requires less memory at the risk of more false positives. The false positive rate can be computed as~\cite{bloom-filter}:  $p \approx (1 - e^{\frac{N*h}{ |BF|}})^{h}$; where $h$ is the number of hash functions and $N$ is the number of data items inserted to the Bloom filter.
For a given $ |BF|$ and $N$, the value of $h$ that minimizes the false positive probability~\cite{bloom-filter, scalable-bloomfilter} is $h = \frac{ |BF|}{N}*\ln{2}$.  
Therefore, we have:  $p \approx (1 - e^{\ln{2}})^{\frac{ |BF|}{N}*\ln{2}}$ which can be simplified to: $\ln{p} = - \frac{ |BF|}{N}*(\ln{2})^2$. Thus $ |BF|$ can be computed as follows:

\begin{equation}
 |BF| = - \frac{N*\ln{p}}{(\ln{2})^2}
\end{equation}

In our design, we select $N = |R_n|$; where $|R_n|$ is the size of the largest input dataset.


\begin{figure}[t]
\centering
\includegraphics [width=0.36\textwidth]{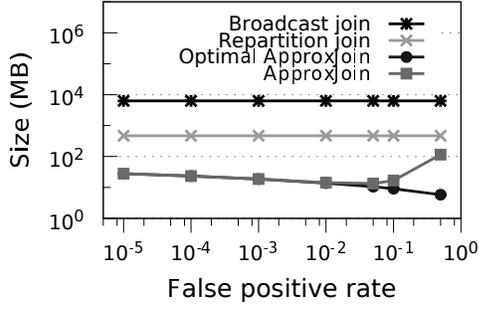}
\caption{Volumes of shuffled data in broadcast join, repartition join, optimal \projecttitle and \projecttitle. Optimal \projecttitle is the case that there are no false positives during the filtering operation.}
\label{fig:simulation-shufflesize}
\end{figure}

We use a simulation-driven approach, based on the aforementioned model, to analyze the trade-off between reducing the shuffled data volume and the desired false positive value in the Bloom filters. We conduct an experiment by using the simulation. We create three input datasets  $R_1$, $R_2$, and $R_3$; where $|R_1| = 10000$, $|R_2| = 1000000$, $|R_3| = 10000000$. We set the overlap fraction to $1\%$; and the number of keys in $R_1$, $R_2$, and $R_3$  to $1000$, $100000$, and $1000000$, respectively. The value of each data item in the datasets follows Poisson distribution with lambda parameter  $10$.
Finally, we set the number of nodes in the cluster $ k = 100$.  
We run the simulation with the input parameters and analyze the shuffled data volume with different false positive values. Figure~\ref{fig:simulation-shufflesize} shows the shuffled data volume of broadcast join, repartition join, optimal \projecttitle, and \projecttitle. Optimal \projecttitle is the case when there are no false positives during the join operation of \projecttitle.
When the false positive rate is set to less than or equal to $0.01$, \projecttitle reaches the optimal case.

This simulation allows us to quickly set the desired false positive parameter for \projecttitle with varying input datasets.

\subsection{Computational Complexity}
Since \projecttitle significantly reduces the communication overhead for distributed join operations, it becomes important to ensure that the bottleneck is not shifted to another part of the system, potentially hindering improved performance. Thus,  another important aspect of the performance analysis is the computational complexity, 
which theoretically represents the amount of time required to execute the proposed algorithm. Here, we provide the computational complexity analysis of our sampling mechanism ($\S$\ref{sec:sampling}) in comparison with the broadcast and repartition  join mechanisms.

Consider that we want to perform a join operation for $n$ inputs $R_1 \Join R_2 \Join ... \Join R_n$, where $R_i (i = 1, ..., n)$ is the input dataset. The inputs contain $m$ join keys $C_j (j = 1, ..., m)$. Let $|r_{ij}|$ be the number of data items participating in the join operation from input $R_i$ with join key $C_j$. 

In repartition join or broadcast join, we need to perform the full cross product operation over these data items. As a result, the computational complexity for each join key $C_j$ is O($\prod\limits_{i = 1}^{n}|r_{ij}|$).

On the other hand, in \projecttitle, we perform sampling over the cross product operation. As a result, 
for each join key $C_j$, the sampling mechanism performs $b_j$ random selections on each side of the bipartite graph ($\S$\ref{sec:sampling}), where $b_j$ represents the sample size of join key $C_j$. $b_j$ is computed as $s*\prod\limits_{i = 1}^{n}|r_{ij}|$, where $s$ is the sampling fraction. Thus, the computational complexity of the proposed sampling mechanism is O($b_i$). Rewriting $b_j$ as $s*\prod\limits_{i=1}^{n}|r_{ij}|$, the computational complexity for each join key $C_j$ becomes O($s*\prod\limits_{i=1}^{n}|r_{ij}|$)).
To summarize, the computational complexity of \projecttitle is lower than the complexity of the broadcast and repartition join mechanisms by a factor of $s$.

\section{Bloom Filter Configuration}
\label{sec:discussion}

\begin{figure}[t]
\centering
\includegraphics[width=0.36\textwidth]{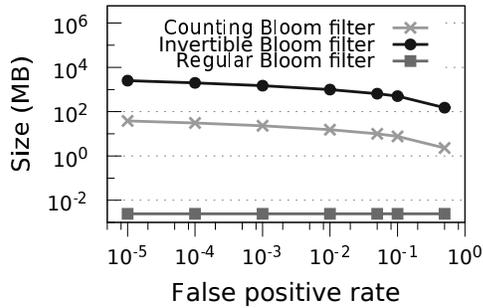}
\caption{Comparison of size of different Bloom filters with varying false positive rates.}
\label{fig:variant-bloomfilter-comparison}
\end{figure}

We discuss three alternative design choices for Bloom filters that we considered in \projecttitle to filter the redundant items (step 1). 
To evaluate different variants of Bloom filters in terms of size and computation cost, we used a simulation with one input dataset containing $100K$ data items and built the corresponding Bloom filters.
Figure~\ref{fig:variant-bloomfilter-comparison} shows the size of each Bloom filter used.

\myparagraph{I: Invertible Bloom filter} 
In addition to the membership check, an {\em Invertible Bloom Filter} (IBF))~\cite{invertable-bloomfilter} also allow to {\tt get} the list of all items present in the filter. As a result, the participating join items can be obtained by using the subtraction operation of the IBF.
However, the IBF comes at a higher cost for computation and storage of the filter: Each cell in an IBF is not a single bit as a regular Bloom filter, but a data structure with a count maintaining the number of collisions and an invertible value of keys. Moreover, just as regular Bloom filters have false positives, there is a probability that a {\tt get} operation returns a ``not found'' result, although the data might still be in the filter, but due to collisions it cannot be found. This probability is the same as the false positive rate for the corresponding Bloom filter. Thus, the filtering step may have {\em false negatives} (due to the ``not found'' result), negatively affecting the join result. Note that such a false negative is not possible with the regular Bloom filters.

\myparagraph{II: Counting Bloom filter} One can also use a {\em Counting Bloom Filter} (CBF)~\cite{counting-bloomfilter} for the filtering stage. CBFs also provide the {\tt remove/subtraction} operation, similar to IBFs, but not the {\tt get} operation. Unlike an IBF, each cell in a CBF is only a count that tracks the number of collisions. As a result, CBFs can be considerably smaller than IBFs (see Figure \ref{fig:variant-bloomfilter-comparison}). However, the size of CBFs is still significantly larger than of regular Bloom filters (see Figure \ref{fig:variant-bloomfilter-comparison}).


\myparagraph{III: Scalable Bloom filter} In our design, we need to know the size of the input datasets for configuring optimal values for the Bloom filters. In practice, however, this information may not always be available in advance. As a result, non-optimal values for Bloom filters may be chosen. To address this problem, we could employ {\em Scalable Bloom filters} (SBFs)~\cite{scalable-bloomfilter}, where the input dataset can be represented without knowing the number of data items to be put in the filter. This mechanism adapts to the growth of the input size by using a series of regular Bloom filters of increasing sizes and tighter error probabilities. 

To build a global SBF (as our join filter), we need to merge local SBFs from all worker nodes in the cluster. Unfortunately, the current design and implementation of SBFs do not support the union operation to perform this merging. We show how to implement this merge operation by creating a pull request\footnote{https://github.com/joseph\-fox/python\-bloomfilter/pull/11} to the SBF repository. Our implementation takes advantage of the fact that SBFs contain a set of regular Bloom filters. As a result, we perform the union operation between two SBFs by executing the union of regular Bloom filters under the hood.

\end{document}